\documentclass[final,5p,times,twocolumn,10pt]{elsarticle}

\usepackage{color}

\usepackage{url}

\usepackage{amssymb}

\journal{Journal of Computational Physics}

\begin{document}

\begin{frontmatter}



\title{Implementation of PDE models of cardiac dynamics on GPUs using OpenCL}


\author[GT]{Christopher D. Marcotte\corref{cor1}}
\ead{cmarcotte3@gatech.edu}
\author[GT]{Roman O. Grigoriev}
\ead{roman.grigoriev@physics.gatech.edu}

\address[GT]{School of Physics, Georgia Institute of Technology, Atlanta, GA 30332-0430, USA}
\cortext[cor1]{Corresponding author.}

\begin{abstract}
Graphical processing units (GPUs) promise to revolutionize scientific computing in the near future. Already, they allow almost real-time integration of simplified numerical models of cardiac tissue dynamics. However, the integration methods that have been developed so far are typically of low order and use single precision arithmetics. In this work, we describe numerical implementation of double precision integrators required by, e.g., matrix-free Newton-Krylov solvers and compare several higher order, fully explicit numerical methods using finite-difference discretization of a range of models of two-dimensional cardiac tissue.
\end{abstract}


\end{frontmatter}



\section{Introduction}

Graphical processing units (GPUs) are becoming increasingly popular in numerical simulations of partial differential equations (PDEs) arising in different areas of science and engineering. Numerical integration of reaction-diffusion-like PDEs describing the dynamics of cardiac tissue is an example of an application where GPUs have been used successfully to achieve considerable speed-up compared to execution on conventional central processing units (CPUs), at comparable hardware cost and complexity. The speed-up is achieved by using numerical algorithms which take advantage of GPUs' parallel architecture which includes a few hundreds to a few thousands of individual processing units highly optimized for algebraic operations.

While the GPU hardware has evolved at an astonishing pace over the last decade, numerical methods that take advantage of this architecture are lagging behind. Traditionally, the discretized problem to be solved was recast into a graphical context to compute image-state values (e.g., color) using shader functions. This allowed numerical integrators to be written in languages developed for 3D computer graphics, modifying textures and vertices of triangle meshes as a simple analogy to a physical model. This approach, wrought with difficulties and limitations, has been superceded in the more recent past by the development of general-purpose computation-oriented languages such as CUDA
and OpenCL,
developed for the highly parallel architectures of modern GPUs in a framework more akin to the Message Passing Interface (MPI),
which has become a standard platform for high performance computing.

CUDA is a proprietary language specific to Nvidia GPU devices. OpenCL is a more general language which can target computing accelerators of many kinds, including, but not limited to, GPUs and CPUs. Both languages are subsets of C99 and possess different advantages: CUDA allows the use of a number of Nvidia-optimized libraries for common mathematical operations, while OpenCL delivers source code portability and the ability to target heterogeneous computational resources. Both software platforms support multiple levels of discrete granularity, from global, to local, to thread level parallelism. 

While single-precision arithmetics might be sufficient for the forward time-integration of general PDEs, many problems require double-precision calculations. For instance, chaotic solutions of PDEs (e.g., turbulence in the Navier-Stokes equation or arrhythmic dynamics in cardiac tissue models) characterized by positive Lyapunov exponents become unreliable on very short time scales, when computed in single precision, regardless of the discretization scheme or time step. Another example is matrix-free calculations, such as the Newton-Krylov method which requires double precision to accurately compute the Jacobian of finite-time flows by finite-differencing nearby solutions. Existing single-precision GPU codes generically rely on features (shaders, texture memory) which do not allow straightforward generalization to double-precision calculations. Consequently, double-precision integrators have to be developed and implemented from scratch.

Our objective here is to show how double-precision numerical integrators for reaction-diffusion type PDEs can be implemented using OpenCL. We will illustrate this using several models of cardiac tissue in two dimensions (although the approach itself is not limited to 2D). The paper is organized as follows. We describe the cardiac models in Sect. \ref{s:cardio} and the numerical schemes in Sect. \ref{s:numer}. Sect. \ref{s:implem} discusses the general constraints of GPU computing, and how they affect the implementation via OpenCL.
The numerical results comparing different models and different methods are presented in Sect. \ref{s:results} and possible avenues for further development are discussed in Sect. \ref{s:optim}. Finally, Sect. \ref{s:conclus} presents our conclusions.

\section{Models of Cardiac Tissue\label{s:cardio}}

Dynamics of cardiac tissue are typically modeled using systems of coupled PDEs of reaction-diffusion type 
\begin{equation}\label{eqn:rdeqn}
  \partial_{\tau}\mathbf{z} = \nabla \cdot \hat{\mathcal{D}} \nabla \mathbf{z} + \mathbf{N}[\mathbf{z}],
\end{equation}
where $\mathbf{N}[\mathbf{z}]$ is a nonlinear function of $\mathbf{z}$ and $\hat{\mathcal{D}}$ is a diffusion tensor which describes 
interaction between neighboring cardiac cells. The variable $\mathbf{z} = (u, \mathbf{h})^{T}$ is a vector field which includes the voltage variable $u$ and a number of gating variables, $\mathbf{h}$. Typically, the diffusion tensor $\hat{\mathcal{D}}$ is determined by the electrical conductivity of the tissue, across and along the muscle fibers, due to the gap junctions between adjacent cardiac cells (cardiomyocytes). In this work we follow the common assumption that the tissue is isotropic and homogeneous, so that $\hat{\mathcal{D}}$ reduces to a matrix with scaled unit blocks on the diagonal.

We will compare three different models: Fitzhugh-Nagumo \cite{FitzH61}, Karma \cite{Karma1994}, and Bueno-Orovio-Cherry-Fenton \cite{Bueno2008}.
In each case we will use Neumann boundary conditions, often referred to as a ``no-flux'' boundary condition in the cardiac literature, which represent, e.g., the vanishing of the electrical current at the boundary of the tissue
\begin{equation}\label{eqn:BC:N}
 \mathbf{n} \cdot \nabla \mathbf{z}(\mathbf{r}) = 0, \qquad
\mathbf{r} \in \partial\Omega,
\end{equation}
where $\mathbf{n}$ is the unit normal to the boundary.

The Fitzhugh-Nagumo system
\begin{eqnarray}\label{eqn:fhn}
  \partial_{t}u\!\!\!\! &=&\!\!\!\! \,\,\,\nabla^{2} u + u - v - u^{3} \nonumber \\
  \partial_{t}v\!\!\!\! &=&\!\!\!\! \delta\nabla^{2}v + \varepsilon(u - a_{1}v - a_{0})
\end{eqnarray}
is a canonical model of excitable media originally derived for neural rather than cardiac tissue, but shares many features with cardiac models. Neurons, like cardiomyocytes, are coupled electrically, although neurons also exhibit long-range coupling via axons and dendritic trees, explaining the second diffusion constant $\delta$. 
Just like most cardiac models, Fitzhugh-Nagumo possesses spiral wave solutions for an appropriate choice of parameters. 
In this work we used nondimensional parameters with values $(\delta, \varepsilon, a_{1}, a_{0}) = (1.5, 0.05, 1.5, -0.1)$ and spatial discretization $\delta x =\delta y = 0.04$.

The two-variable Karma model
\begin{eqnarray} \label{eqn:karma}
 \partial_{\tau}u\!\!\!\! &=&\!\!\!\! \gamma\nabla^{2} u + \tau_{u}^{-1}\left( \frac{u^{2}}{2}[1 - \tanh(u - u_{h})][u^{*} - v^{M}] - u \right) \nonumber \\
 \partial_{\tau}v\!\!\!\! &=&\!\!\!\! \tau_{v}^{-1}\left( \left[\frac{1}{1 - e^{-Re}} - v \right] H_{k}(u - u_{v}) - v H_{k}(u_{v} - u) \right)
\end{eqnarray}
also possesses spiral wave solutions, but these solutions can exhibit alternans -- an instability manifested in the variation of the wavelength which can initiate spiral wave break-up. This makes the Karma model of greater relevance for cardiac dynamics although it is only marginally more complex than Fitzhugh-Nagumo.
We use parameter values from Ref.~\cite{Karma1994}: $(Re, M, \gamma) = (1.0, 6, 0.0011)$ and spatial discretization  $\delta x =\delta y = 0.0262$ cm.

The four-variable Bueno-Orovio model 
\begin{eqnarray} \label{eqn:bochfe}
 \partial_{t}u\!\!\!\! &=&\!\!\!\! \mathcal{D} \nabla^{2}u - (J_{fi} + J_{so} + J_{si}) \nonumber \\
 \partial_{t}v\!\!\!\!	&=&\!\!\!\! \alpha_{v}(u) v + \beta_{v}(u) \nonumber \\
 \partial_{t}w\!\!\!\!	&=&\!\!\!\! \alpha_{w}(u) w + \beta_{w}(u) \nonumber \\
 \partial_{t}s\!\!\!\!	&=&\!\!\!\! \left(\frac{1 + \tanh(k_{s}(u - u_s))}{2} - s\right)\tau_{s}(u)
\end{eqnarray}
was designed as a ``minimal'' model capable of accurately reproducing the shape, propagation speed, and stability of spiral waves predicted by detailed ionic models of cardiac tissue which contain tens of variables.
In this work the epicardial parameter set from Ref.~\cite{Bueno2008}, with $\mathcal{D} = 1.171$ cm$^{2}/$s and discretization $\delta x = \delta y = 0.03$ cm, is used.

The electrical currents $J_{fi}$, $J_{so}$, and $J_{si}$ in (\ref{eqn:bochfe})
\begin{eqnarray*}\label{eqn:bochfe:currents}
 J_{fi}\!\!\!\! &=&\!\!\!\! -v H_{k}(u - \theta_{v}) (u - \theta_{v}) (u_{u} - u)/\tau_{fi} \\
 J_{so}\!\!\!\! &=&\!\!\!\! (u - u_{o}) H_{k}(\theta_{w} - u)/\tau_{o}(u) + H_{k}(u - \theta_{w})/\tau_{so}(u) \\
 J_{si}\!\!\!\! &=&\!\!\!\! -H_{k}(u - \theta_{w}) w s/\tau_{si}(u) \\
\label{eqn:bochfe:greek}
 \alpha_{x}(u)\!\!\!\! &=&\!\!\!\! -H_{k}(u - \theta_{x})/\tau_{x}^{+} - H_{k}(\theta_{x} - u)/\tau_{x}^{-}\\
 \beta_{x}(u)\!\!\!\! &=&\!\!\!\! H_{k}(\theta_{x} - u) x_{\infty}/\tau_{x}^{-}
\end{eqnarray*}
represent the flux of ions (calcium, potassium, and sodium, primarily) through ionic channels in the cellular membrane, whereas the diffusive term represents inter-cellular current flow through gap junctions between neighboring cells.
The gating variable dynamics switch between three characteristic time scales per variable. Two time scales are associated with the relaxation dynamics, and only one with the excitation of individual cells. Physiologically, they describe the opening and closing of the ionic channels.

Traditionally, $H_{k}(x)$ is taken to be the Heaviside function $\Theta(x)$ (a discontinuous step function centered at the origin). However, the discontinuity is both unphysical and unsuitable for calculations involving derivatives of the solution (such as the Newton-Krylov methods). Canonical choices for a smoothed step function such as $H_{k}(x) = 0.5(1 + \tanh(kx))$ are expensive to compute on a GPU, so we chose instead a smoothed version of the step function with cubic interpolation
\begin{displaymath} \label{eqn:step}
   H_{k}(x) = \left\{
     \begin{array}{ll}
       0, & kx<-0.5, \\
       (3 - 2 k x - 1)(k x + 0.5)^2, & -0.5 \leq kx \leq 0.5, \\
       1, & kx > 0.5
     \end{array}
   \right.
\end{displaymath}
which is much faster to compute and approaches the Heaviside function as parameter $k$, which controls the width $w=1/k$ of the transition region, goes to infinity. In this work we used $k=28.4$ which strikes a good balance between differentiability and a small width $w\approx 0.035$ of the transition region. For comparison, the range of the gating variables is $(0, 1)$ and the range of the voltage is $(-1, 1)$, $(0, 3.5)$, and $(0, 1.55)$ for the Fitzhugh-Nagumo, Karma, and Bueno-Orovio models, respectively.

\section{Numerical Model\label{s:numer}}

For each of the three models, the physical domain was spatially discretized using finite differences on an $N\times N$ Cartesian grid with spacing $\delta x$ and $\delta y$. The continuous time variable was discretized by a fixed time-step, $\delta t$. The diffusion operator, $\mathcal{D}\nabla^{2}$, was approximated by a nine-point stencil over the nearest neighbors and nearest neighbors along the diagonal
\begin{equation}
\label{eqn:diffusion}
 (\delta x \, \delta y)\nabla^{2}U_{i\,j} = \sum_{k=-1}^{1} \sum_{l=-1}^{1} \alpha_{k\,l}U_{i+k\,j+l},
\end{equation}
where the diagonal terms are uniformly weighted ($\alpha_{\pm\,\pm} = \alpha_{d}$), the axial terms are weighted equivalently ($\alpha_{\pm\,0}=\alpha_{0\,\pm}=\alpha_{a}$), and the central point is weighted such that the stencil is balanced ($\alpha_{00} + 4(\alpha_{d} + \alpha_{a}) = 0$).

The specific form of the stencil is the result of symmetry considerations. As shown in \cite{Wardetzky:2007:DLO:1281991.1281995}, the discrete Laplace operator may not satisfy all the canonical properties of the continuous operator. In this work we sought to preserve the locality property of the continuous operator (because of the association with diffusion processes), linearity, and the continuous symmetries of the underlying dynamical equations: translational and rotational symmetry. Computational restrictions preclude the implementation of discrete operators which preserve rotational symmetry to arbitrarily high-order, but the stencil with diagonal elements respects rotational symmetry to a far greater degree than the canonical central-difference discrete operator.

\begin{figure}
 \begin{center}
  \includegraphics[width=0.2\textwidth]{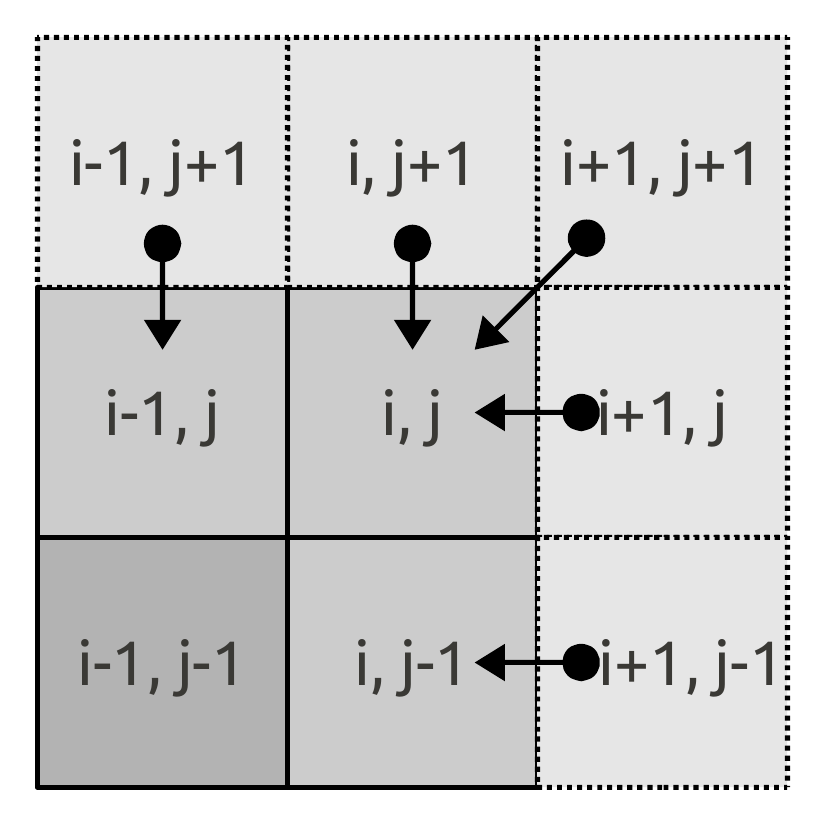}
 \end{center}
 \caption{Diagram of the ghost-point method of enforcing Neumann boundary conditions. Locations outside the domain (light gray) are assigned boundary values (medium gray), effectively mirroring the value of the diffusive field. The dark gray cell is in the interior of the domain.}
 \label{fig:ghost}
 \end{figure}

The implementation of the Neumann boundary conditions~(\ref{eqn:BC:N}) relies on the calculation of the Laplacian along the boundary of the domain. For each discrete spatial displacement $(k,l)$ relative to a boundary grid node with spatial index $(i,j)$, logical indexing is used to flip displacement vectors which cross the domain boundary back into the domain. This procedure is illustrated in Fig.~\ref{fig:ghost}. The primary benefit of this method is simplicity, and thus, extensibility, of the implementation. Larger stencils can be implemented with no changes to the boundary condition code, only to the Laplacian calculation. Periodic boundary conditions can be imposed instead by mapping to the opposite-edge values of the domain, as opposed to adjacent-edge values. Due to the nature of the GPU architecture, branching threads incur some performance penalty. Pre-populating the boundaries of the domain using canonical subregion techniques was found to not be appreciably faster or slower than logical indexation, but for fixed-grid sizes, the former must yield a smaller computational domain.

The time-integration methods explored in this work include implicit and explicit methods, with linear and exponential time differencing steps, and first to fourth order theoretical convergence rates. They include the Euler, Implicit Euler, Rush-Larsen, Heun, and classic Runge-Kutta methods of integration. (Implicit Euler and Rush-Larsen were implemented only for the Bueno-Orovio model.) We should point out that the use of fully explicit methods over implicit, or semi-implicit, methods allows one to take advantage of the native implementation of fused multiply-add operations (over division which incurs a performance penalty). All methods were implemented using a fixed time-step for the purposes of convergence analysis.

\section{Implementation on the GPUs\label{s:implem}}

All the results presented in this paper were produced using an Nvidia GeForce GTX 680 GPU which uses ``Kepler'' architecture. The card was installed in a workstation with an Intel Core i7-3770K CPU clocked at 3.50GHz and 16GB system memory. This GPU is capable of double precision arithmetic at $1/24$th the computational performance relative to the single precision computation power of 3090.4 Gflop/s. Thus, the peak theoretical power of the GPU is roughly 128 Gflop/s in double precision. The peak memory bandwidth is 192 GB/s, with a core clock frequency of 1006 MHz, 48 KBytes of L2 cache (local memory), 2048 MBytes of L3 cache (global memory), and 1536 streaming processors (CUDA cores).

Our implementation of the integrators is based on the OpenCL Toolbox-- an open-source bridge between OpenCL and Matlab \cite{oclToolbox}. The software is provided under a generous MIT license, and enables the use of Matlab to execute a host code which invokes OpenCL kernels that advance the solution forward in time. Despite certain limitations (such as the lack of vector data type buffers) the toolbox greatly simplifies the integration between Matlab and OpenCL by abstracting memory management.

Spatial extent and dimensionality of the physical domain as well as the addressing of computational cells within the domain all affect the implementation of the numerical methods on specific hardware. As a reflection of their original purpose, GPUs have an architecture with a natural three-dimensional basis, where each computational thread is indexed by a triplet ${\bf i}=(i_1,i_2,i_3)$, where $0\le i_k<N_k$, $k=1,2,3$. OpenCL has several functions for using indexation to address both global and local memory from within each thread. In particular, the function {\tt global\_work\_size($k$)} returns an integer $N_k$ which defines the size of the computational domain along the $k^\mathrm{th}$-dimension. Similarly, the function {\tt get\_global\_id($k$)} returns the index $i_k$ of the thread along the $k^\mathrm{th}$-dimension.
These two functions describe the smallest-scale discretization of the OpenCL environment, and in this work provide a one-to-one mapping from the discretized spatial domain to the computational array-index. The two-dimensional nature of the calculations presented in this work requires $N_1=N_2=N$ and $N_3=1$, so that $0\le i_1,i_2< N$ and $i_3=0$.

The non-locality of PDEs requires that updates at every point in the computational domain be temporally correlated. For methods which update different spatial locations in parallel, there is no guarantee of simultaneity even between neighbors. Our integration method exploits the massive parallelism of the GPU by assigning unique discrete spatial locations to individual threads by the index mapping defined by the {\tt get\_global\_id()} function. As the thread count vastly outnumbers the number of available stream processors, threads are executed in parallel within 32-thread work-groups as each work-group is passed to the available multiprocessors on the GPU. The execution of the work-groups is serial, and thus complicates the temporal correlation for extended systems. In contrast, systems of Ordinary Differential Equations (ODEs) which possess only local terms require no synchronization. For this reason, our integration method calculates the forward-time updates in lock-step, enforcing synchrony across the domain with blocking read and write operations to global memory. For this reason, the calculation of the action of the Laplace operator in each of the models (\ref{eqn:fhn}), (\ref{eqn:karma}), and (\ref{eqn:bochfe}) is treated separately from the nonlinear, local terms, which can be updated asynchronously.

\begin{figure}[t]
 \begin{center}
 \includegraphics[width=0.46\textwidth]{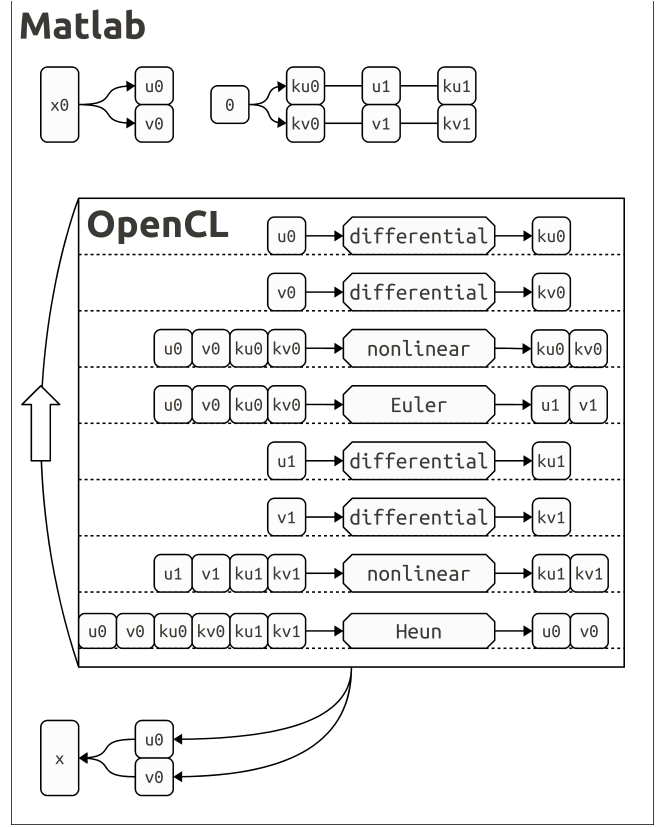}
 \end{center}
 \caption{Block diagram of the Heun integrator function for the Fitzhugh-Nagumo system, detailing the placement of kernel functions and memory objects within the code. The rounded squares represent memory objects, created within the OpenCL context and filled with initial values from the Matlab environment. The octagonal blocks represent OpenCL kernel functions, labeled by their action. They accept the memory objects as inputs (incoming arrows) and write to memory objects, which we will non-specifically refer to as outputs (outgoing arrows). The horizontal dotted lines ($\cdots$) represent blocking write actions to global memory on the GPU which maintain synchronization between threads. Note the loop structure inherent to the OpenCL block, and the eventual return of the state variables \texttt{u0} and \texttt{v0} to the Matlab context.}
 \label{fig:blockDiagram}
 \end{figure}

The general pattern for explicit higher order integrators involves repeated or nested Euler updates, with an $n^\mathrm{th}$ order method requiring at least $n$ updates. In each of the Euler updates the nonlocal (differential) terms on the right hand side are computed first, followed by the calculation of the local (nonlinear) terms, after which the results are combined and used to update each field. The implementation of this procedure is illustrated in Fig.~\ref{fig:blockDiagram} for the Heun method, which represents well the algorithmic structures which arise in higher-order integration methods. 

In general, for an integration method with $n_s$ substeps and a model with $n_v$ fields, out of which $n_d \leq n_v$ involve diffusive coupling, $2(n_s \cdot n_v)$ buffers are needed to store the results of intermediate calculations and the current state of the system. Additionally, $(n_d + 2)\cdot n_s$ kernel function calls per time-step are invoked.

\section{Results and Discussion\label{s:results}}

\begin{figure}[t]
 \begin{center}
  \begin{tabular}{lcl}
 (a)&\hspace{-3mm}
\includegraphics[height=0.24\textwidth]{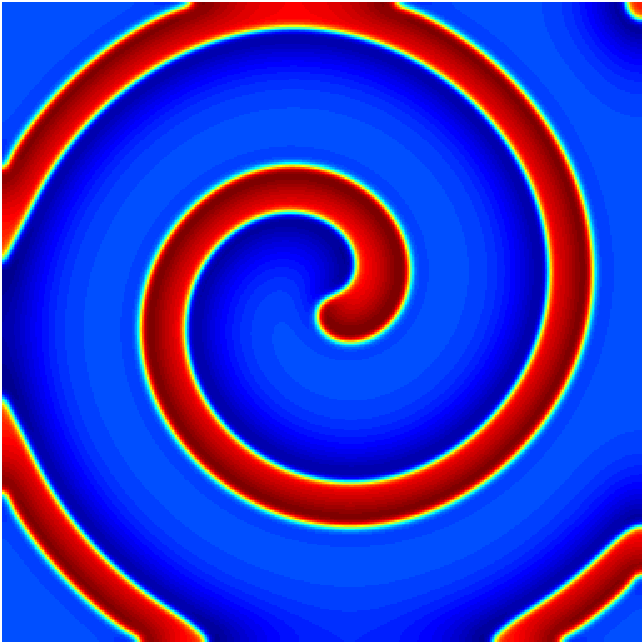}		&
\includegraphics[height=0.24\textwidth]{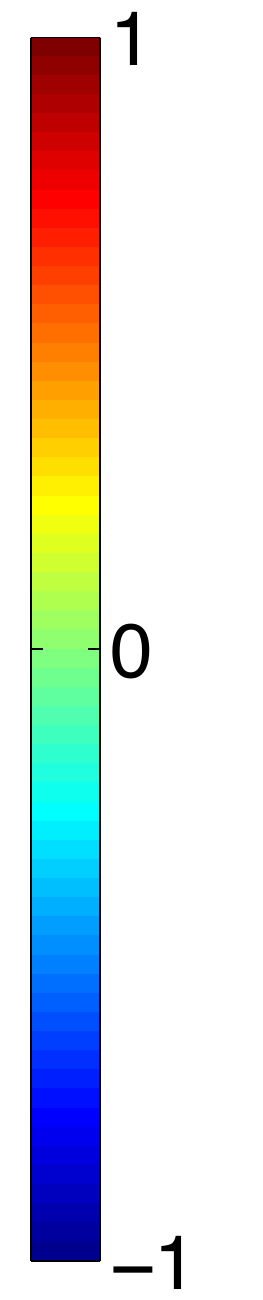}\\
 (b)&\hspace{-3mm}
\includegraphics[height=0.24\textwidth]{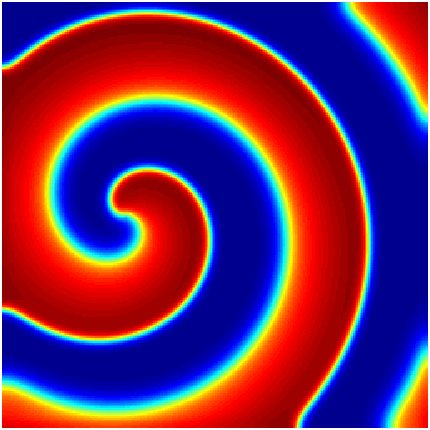}	&\includegraphics[height=0.24\textwidth]{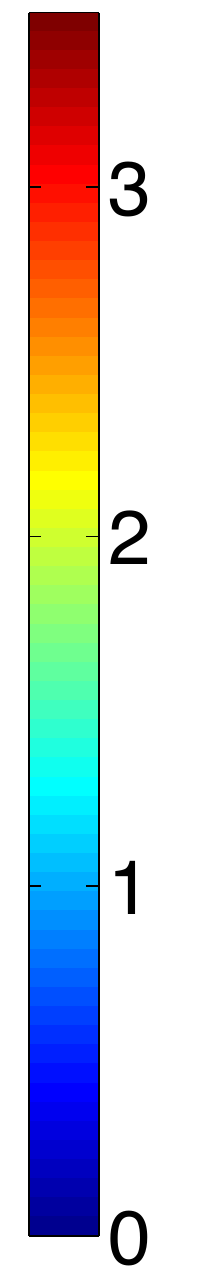}\\
 (c)&\hspace{-3mm}
\includegraphics[height=0.24\textwidth]{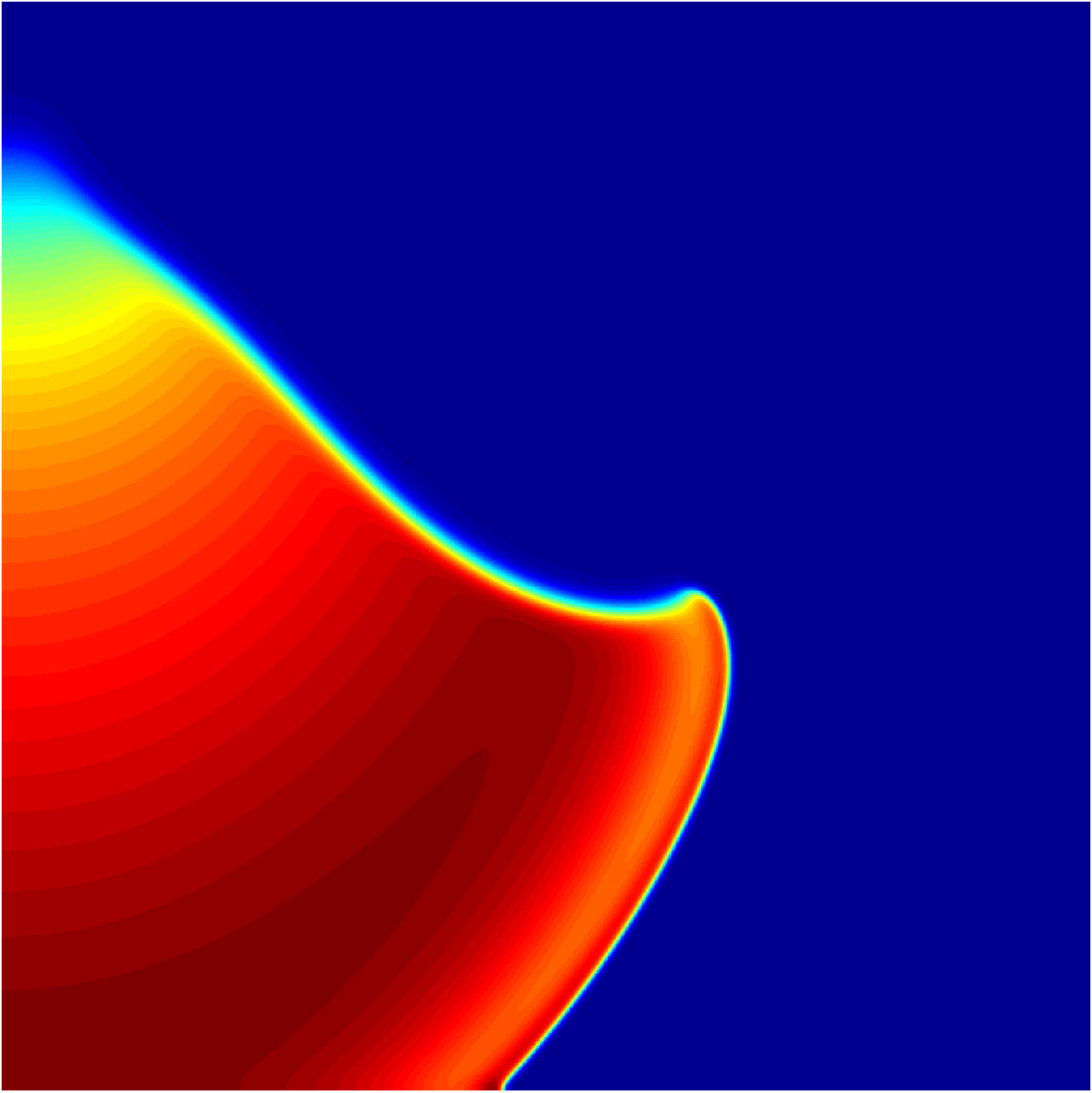}	&\includegraphics[height=0.24\textwidth]{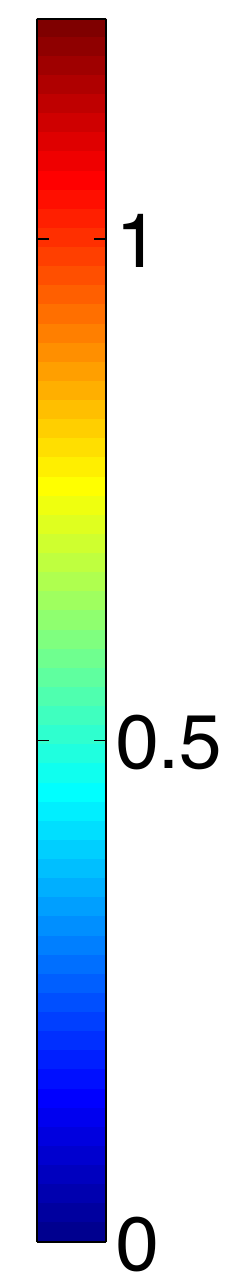} 
  \end{tabular}
 \end{center}
 \caption{Voltage field $u$ for the (a) Fitzhugh-Nagumo, (b) Karma, and (c) Bueno-Orovio models computed on $384\times 384$, $256\times 256$, and $768\times 768$ grids. All three pattern clearly represent spiral waves, although the shape of the wave strongly depends on the model and the choice of parameters. Fringing effects at the boundaries are characteristic of Neumann boundary conditions ${\bf n}\cdot\nabla u=0$.}
 \label{fig:voltage}
 \end{figure}

Typical patterns produced by numerical integration of the three models are presented in Fig. \ref{fig:voltage} which shows snapshots of the voltage field ($u$) computed for the Fitzhugh-Nagumo, Karma, and Bueno-Orovio models. Although all three models produce spiral waves, the shape of the wave and the pitch of the spiral are considerable different. Indeed, only the Bueno-Orovio model produces excitation dynamics that is physiologically relevant.

\begin{figure}
 \begin{center}
  \includegraphics[width=0.48\textwidth]{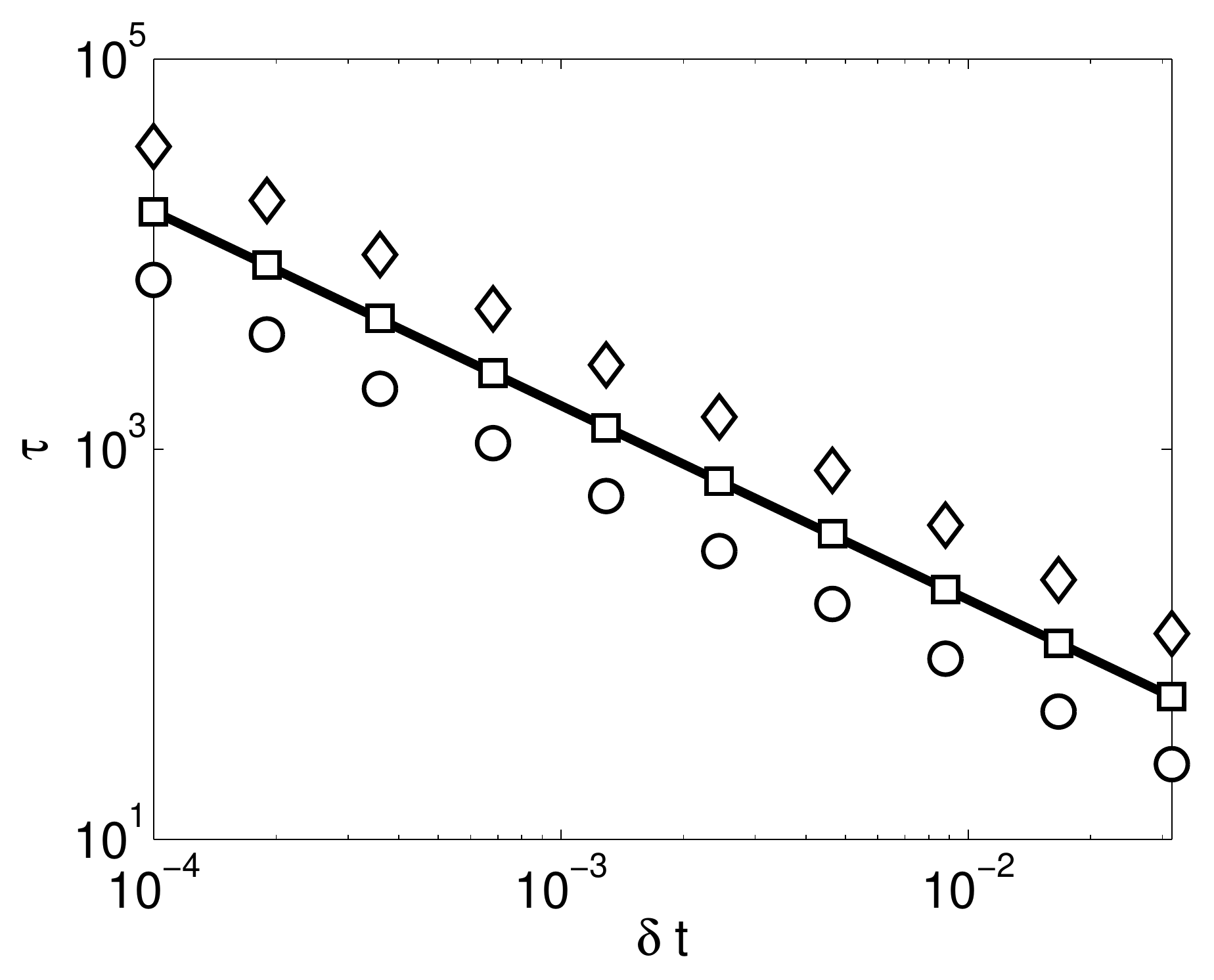} \\
 \end{center}
 \caption{Computation time for the Karma model as a function of time-step $\delta t$. The vertical axis $\tau$ is the wall-time measured by the Matlab timing functions {\tt tic()} and {\tt toc()}, scaled by the integration time interval ($\tau = 1$ is equivalent to a real-time simulation). The solid line represents inverse scaling fit, $\tau\propto\delta t^{-1}$ to the timing results obtained using Heun method. The symbols represent different integration methods summarized in Tab.~\ref{tab:conv}. Calculations were performed on a $256\times 256$ grid.}
 \label{fig:timing}
\end{figure}

In order to quantify and compare the performance of different integrators we computed several metrics. In particular, Fig.~\ref{fig:timing} shows the ratio $\tau$ of the wall clock time required to integrate the Karma model (\ref{eqn:karma}) over a fixed time interval $[0,t]$ and the length $t$ of that interval. The scaling behavior is similar for the Fitzhugh-Nagumo and Bueno-Orovio models (not shown) and shows that the $\tau$ scales as the inverse of the time step $\delta t$ (or linearly with the number of time steps, as expected). It is also proportional to the order of the integration method (or the number of substeps $n_s$), again as expected: the second order Heun method takes approximately twice as long to compute as the Euler method, and the fourth order Runge-Kutta method takes approximately twice as long as the Heun method.

\begin{figure}
 \begin{center}
  \includegraphics[width=0.48\textwidth]{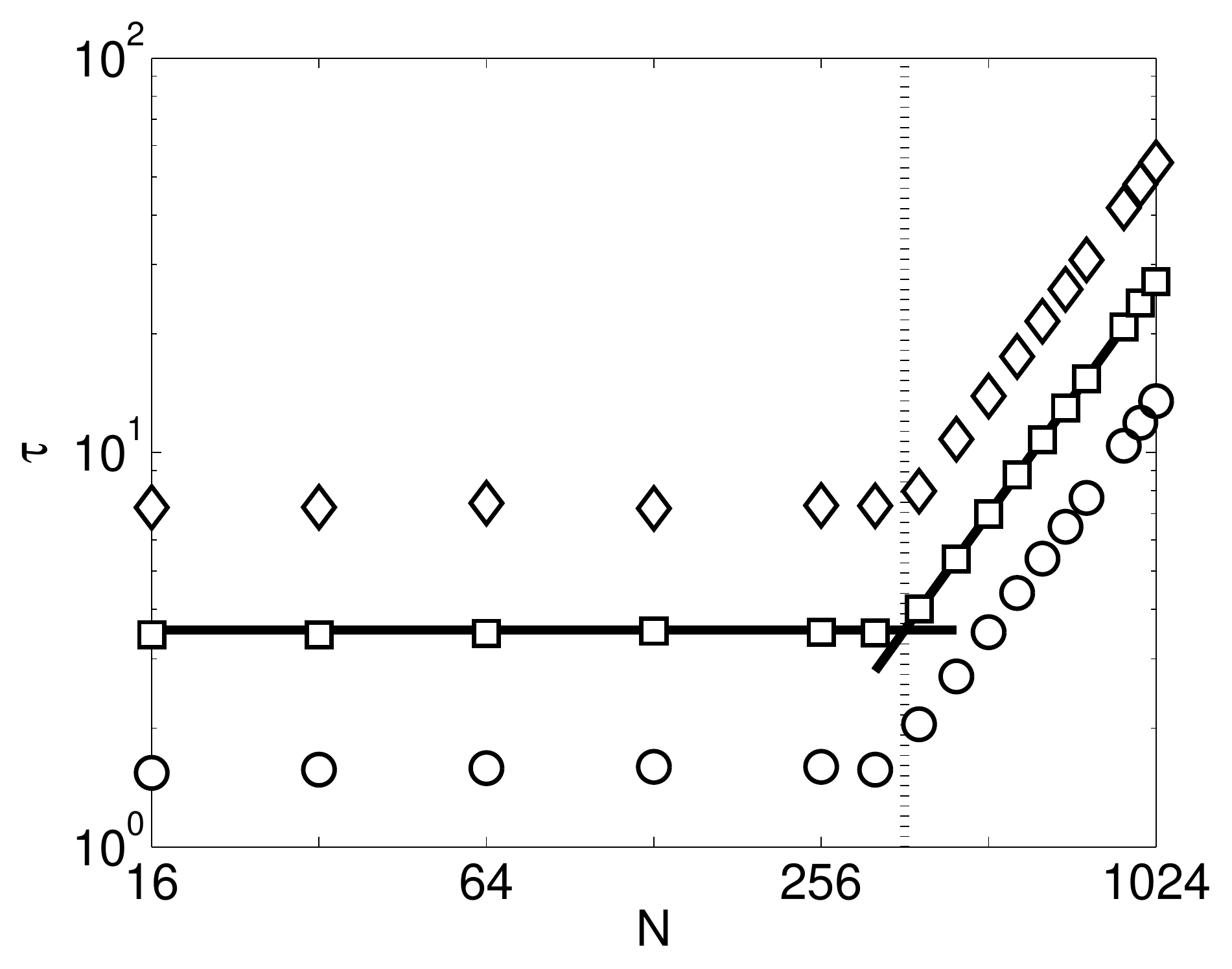} \\
 \end{center}
 \caption{Computation time for the Karma model as a function of the grid size. The solid lines represent a piece-wise linear fit for the Heun method: $\tau=a$ for $N<N_0$ and $\tau=bN^2$ for $N>N_0$. The vertical dotted line corresponds to $N=N_0\approx 362$. Symbol notation is as in Tab.~\ref{tab:conv}. All calculations were performed with $n = \lceil T/\delta t\rceil = 2000$ time-steps, and so the values of $\tau$ in this plot are not directly comparable to those in Fig.~\ref{fig:timing}.}
 \label{fig:grid:all}
\end{figure}

Fig. \ref{fig:grid:all} shows how the wall clock time scales with the size $N$ of the computational grid, also for the Karma model. We find that $\tau$ is effectively independent of $N$ for small grids, and sharply transitions to quadratic scaling in $N$ near $N_0 \approx 362$: $\tau = a\max(1,bN^{2})$, where $b=a/N_0^2$ for the Karma model. This behavior is similar for all methods and models and indicates that for small grids $\tau$ is controlled by the `start up' time required for invoking an OpenCL kernel from within Matlab (e.g., for transferring data between the system memory and the GPU memory), which is independent of the number of forked threads. As the number of threads increases, the start up time is overwhelmed by the time required to execute the threads, which scales as $N^2$. The sharp crossover at $N=N_0$ indicates that execution of the threads and data transfer take place simultaneously, rather than sequentially. Therefore, $N_0$ is dependent on the memory bandwidth \textit{and} the computational power of the GPU.

\begin{figure}
 \begin{center}
 (a)\hspace{-4mm}\includegraphics[width=0.48\textwidth]{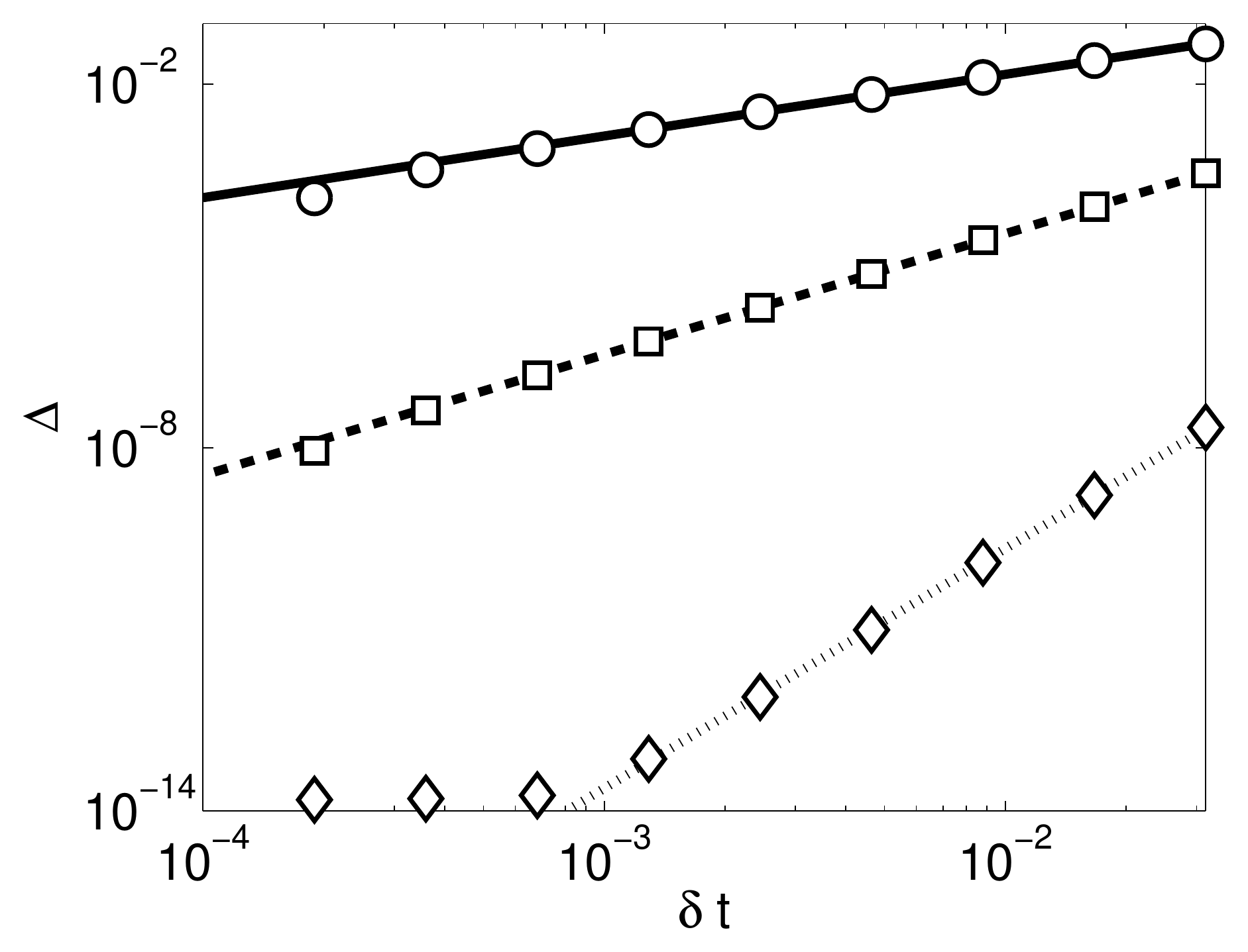} \\
 (b)\hspace{-4mm}\includegraphics[width=0.48\textwidth]{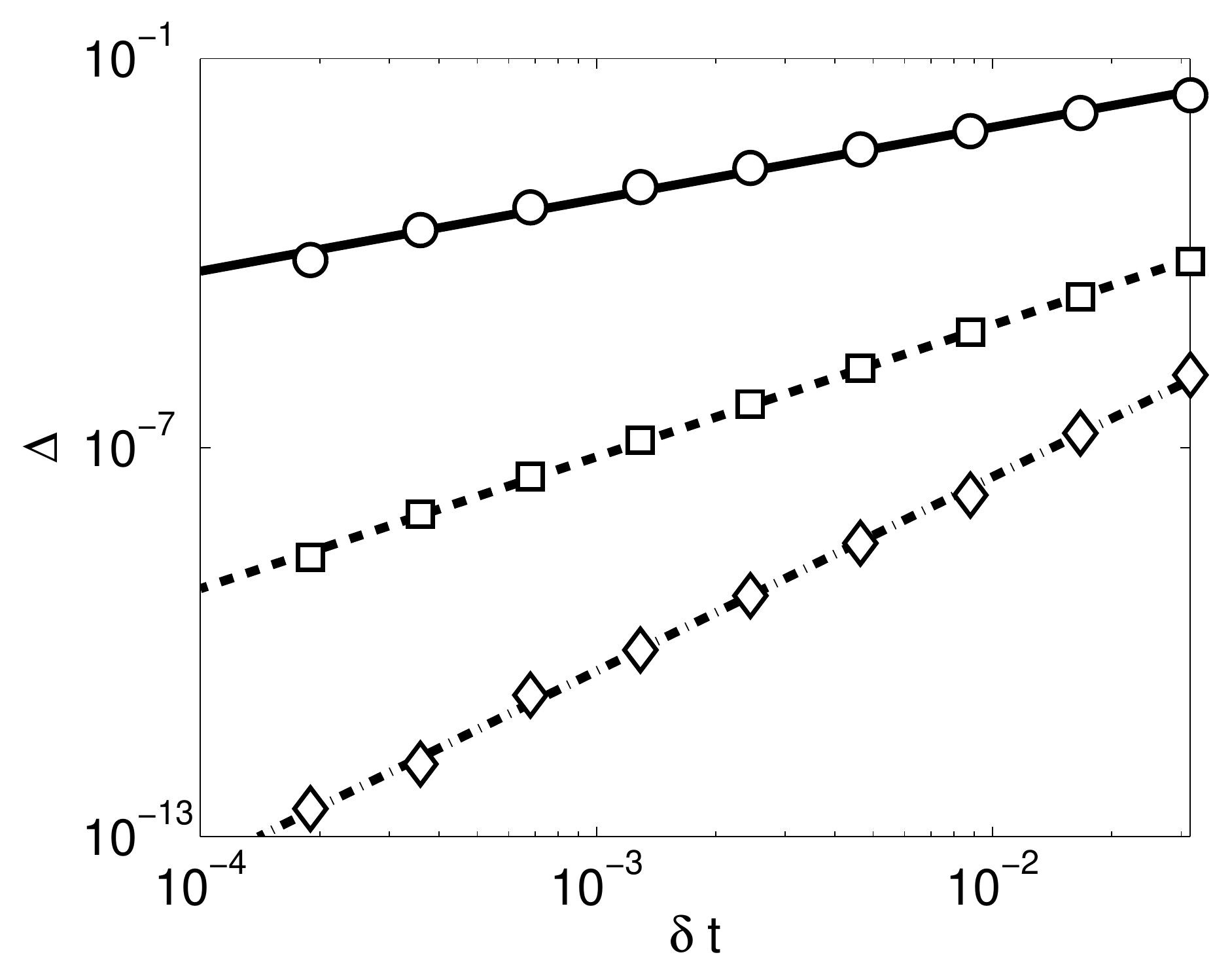} \\
 (c)\hspace{-4mm}\includegraphics[width=0.48\textwidth]{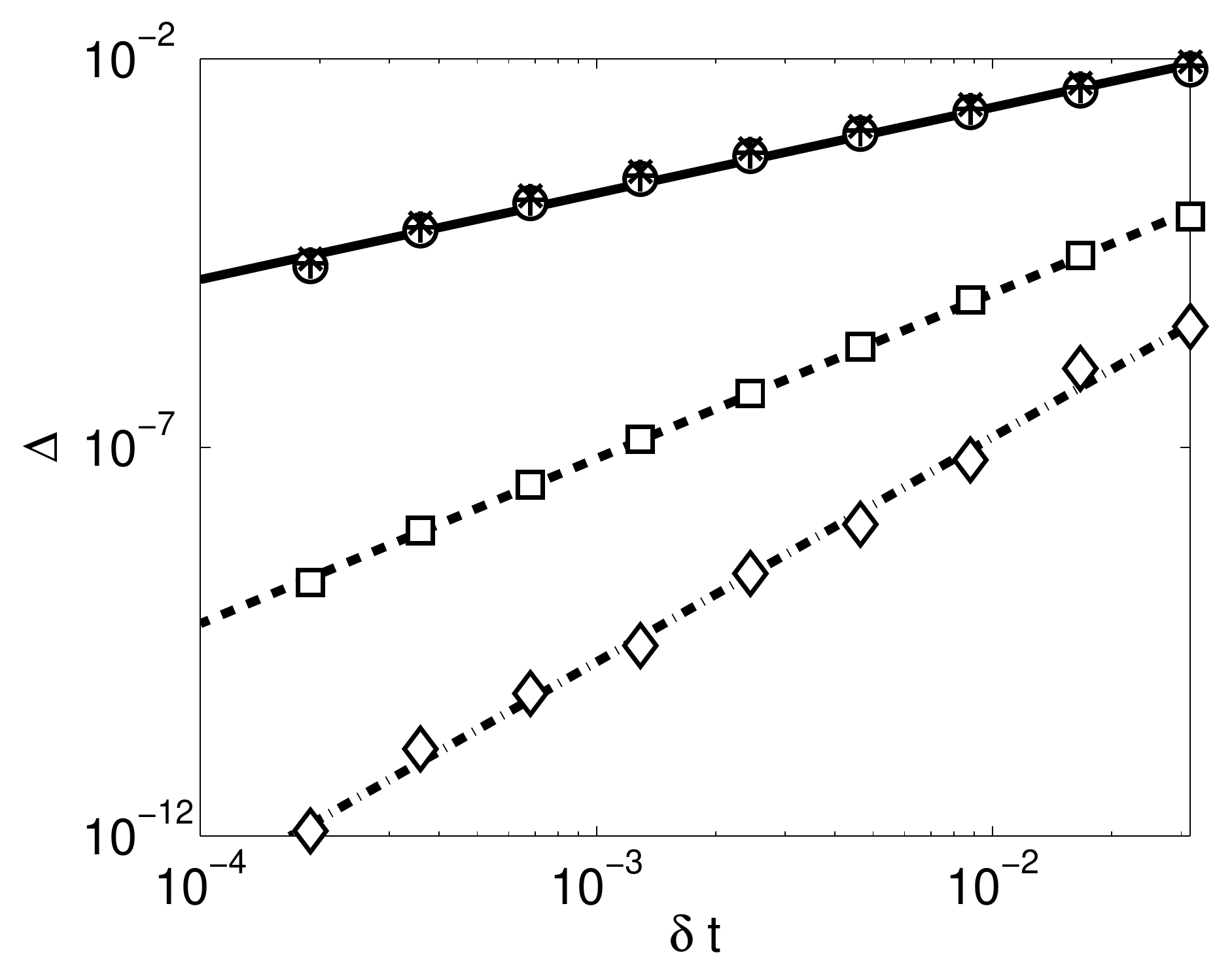}
 \end{center}
 \caption{Accuracy of the integration for (a) the Fitzhugh-Nagumo, (b) Karma, and (c) Bueno-Orovio models as an function of time-step $\delta t$. The vertical axis $\Delta$ is the relative precision measured by the $L_2$-norm difference from a reference solution, scaled by the $L_2$-norm of the reference solution. Symbols designate different integration methods, as in Tab.~\ref{tab:conv}. The solid, dashed, dash-dot, and dotted lines are the power-law fits $\Delta\propto\delta t^{\,\alpha}$ with exponent $\alpha=1$, $\alpha=2$, $\alpha=3$, and $\alpha=4$, respectively. Calculations were performed on a $256\times 256$ grid. The Bueno-Orovio model (c) includes both the Implicit-Euler and Rush-Larsen methods, partially obscured by the Explicit-Euler results.}
 \label{fig:convergence}
\end{figure}

Convergence results for all models and all methods considered here are shown in Fig.~\ref{fig:convergence}. We find that in all cases the accuracy of integration scales as an integral power of the time step $\Delta\propto\delta t^{\,\alpha}$, with the values of exponents $\alpha$ summarized in Tab.~\ref{tab:conv}. In particular, the exponents take the expected values $\alpha=1$ for all first-order methods and $\alpha=2$ for all second-order methods. (The Implicit Euler and Rush-Larsen methods are only applied to the non-diffusive variables, and so the stability-improving properties of the methods are lost, in practice.)

\begin{table}
\begin{center}
\begin{tabular}{lcc|ccc}
Method		&		& Order	& Eq.~(\ref{eqn:fhn})	& Eq.~(\ref{eqn:karma}) & Eq.~(\ref{eqn:bochfe})	\\ \hline
Explicit Euler	& $\circ$ 	& 1 	&	1		&	1		&	1			\\
Implicit Euler 	& $\times$ 	& 1 	&	---		&	---		&	1			\\
Rush-Larsen 	& $+$ 		& 1 	&	---		&	---		&	1			\\
Heun 		& $\square$ 	& 2 	&	2		&	2		&	2			\\
Runge-Kutta 	& $\Diamond$ 	& 4 	&	4		&	3		&	3			\\
\end{tabular}

\end{center}
\caption{\label{tab:conv}Theoretical order of the integration method and the observed scaling exponent $\alpha$ for different models. The values of $\alpha$ have been computed from the best fits shown in Fig. \ref{fig:convergence}. }
\end{table}

In the Fitzhugh-Nagumo model the fourth-order Runge-Kutta method follows theoretical scaling with $\alpha=4$, but exhibits saturation of the relative precision around $10^{-14}$. This is expected as the solution approaches the limit of double precision arithmetic, per element. 

Rather unexpectedly, both Karma and Bueno-Orovio model exhibit cubic rather than quartic convergence for the fourth-order Runge-Kutta method. We have not been able to establish the source of inaccuracies leading to slower than expected convergence rate.

\begin{figure}
 \begin{center}
 (a)\hspace{-4mm}\includegraphics[width=0.48\textwidth]{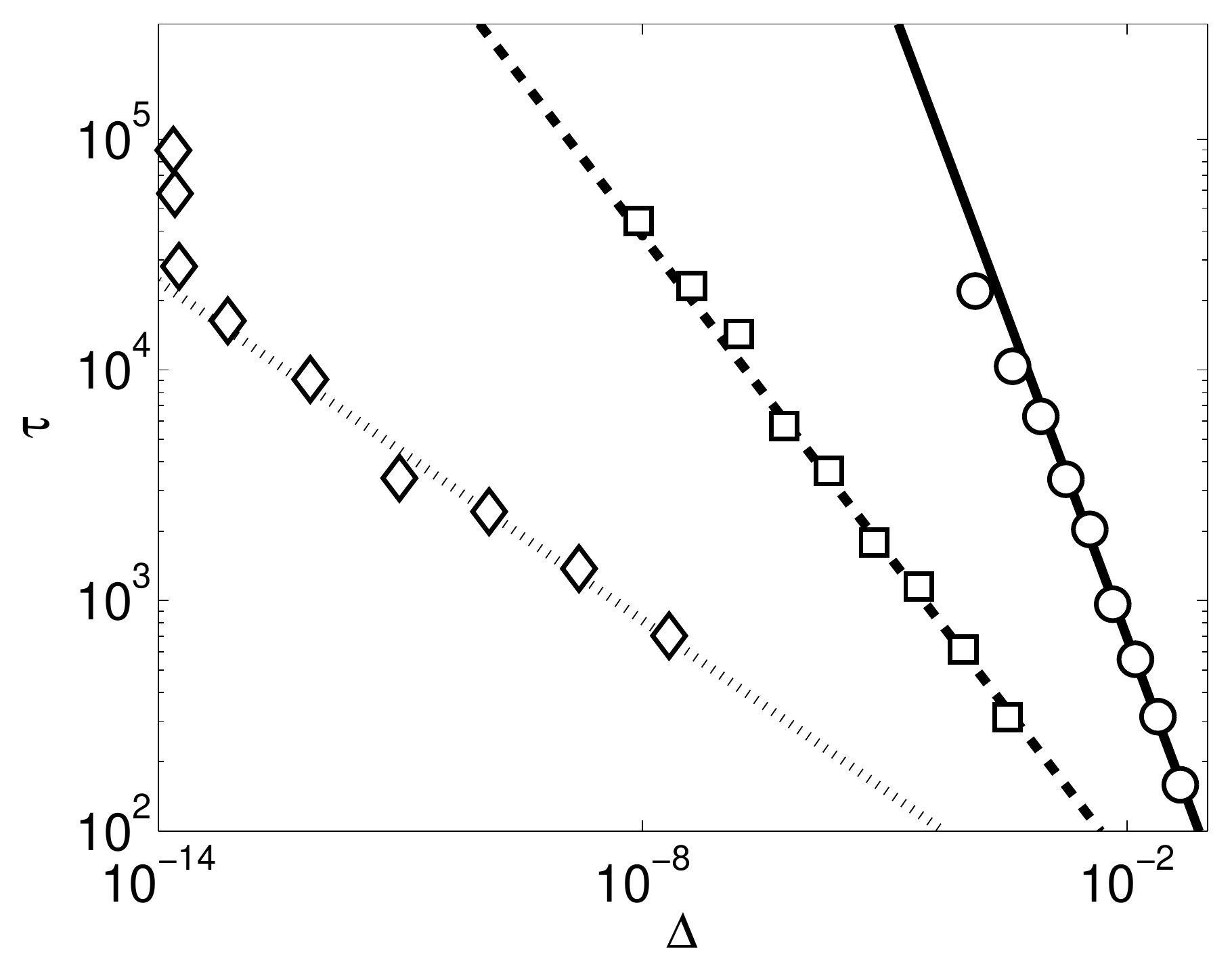} \\
 (b)\hspace{-4mm}\includegraphics[width=0.48\textwidth]{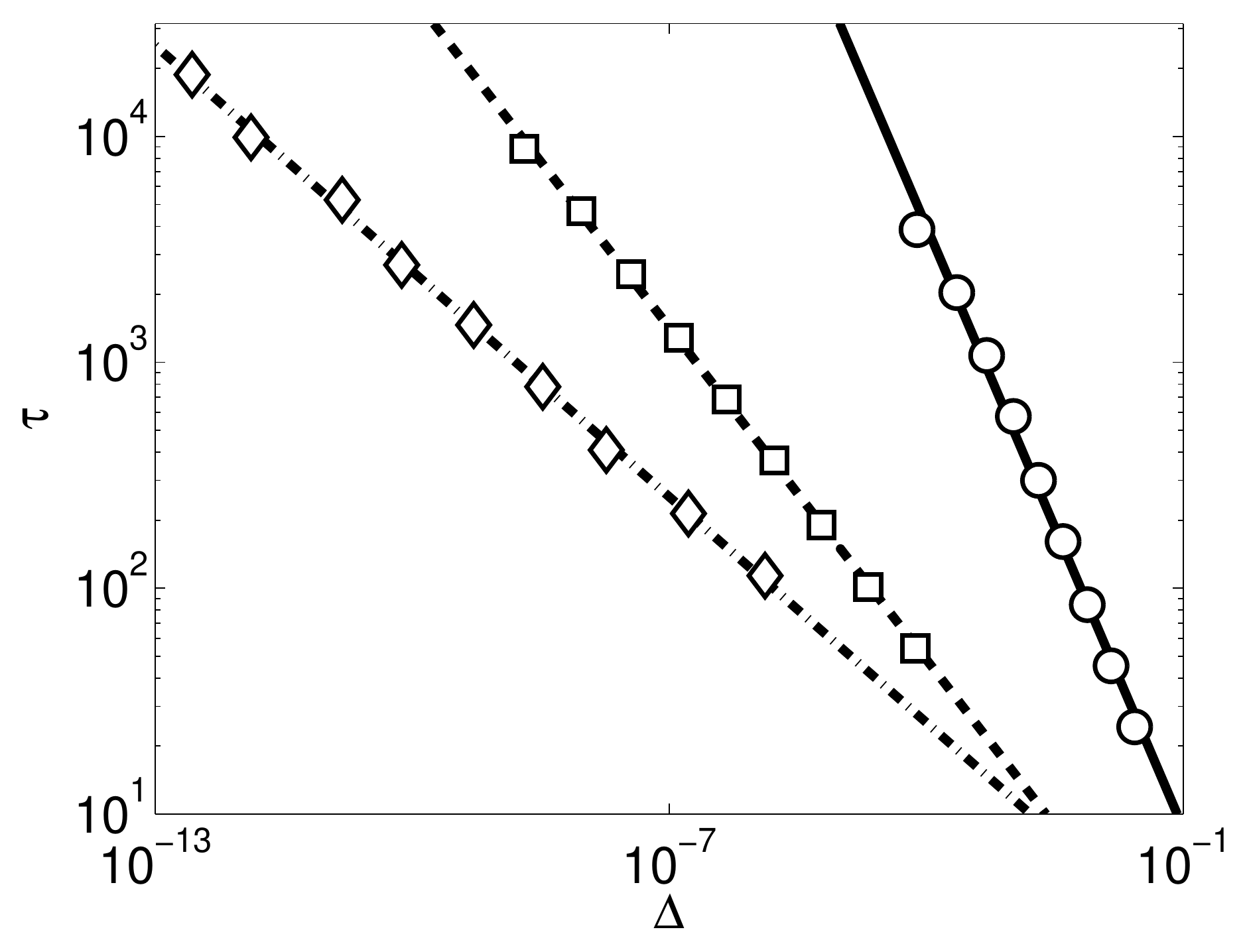} \\
 (c)\hspace{-4mm}\includegraphics[width=0.48\textwidth]{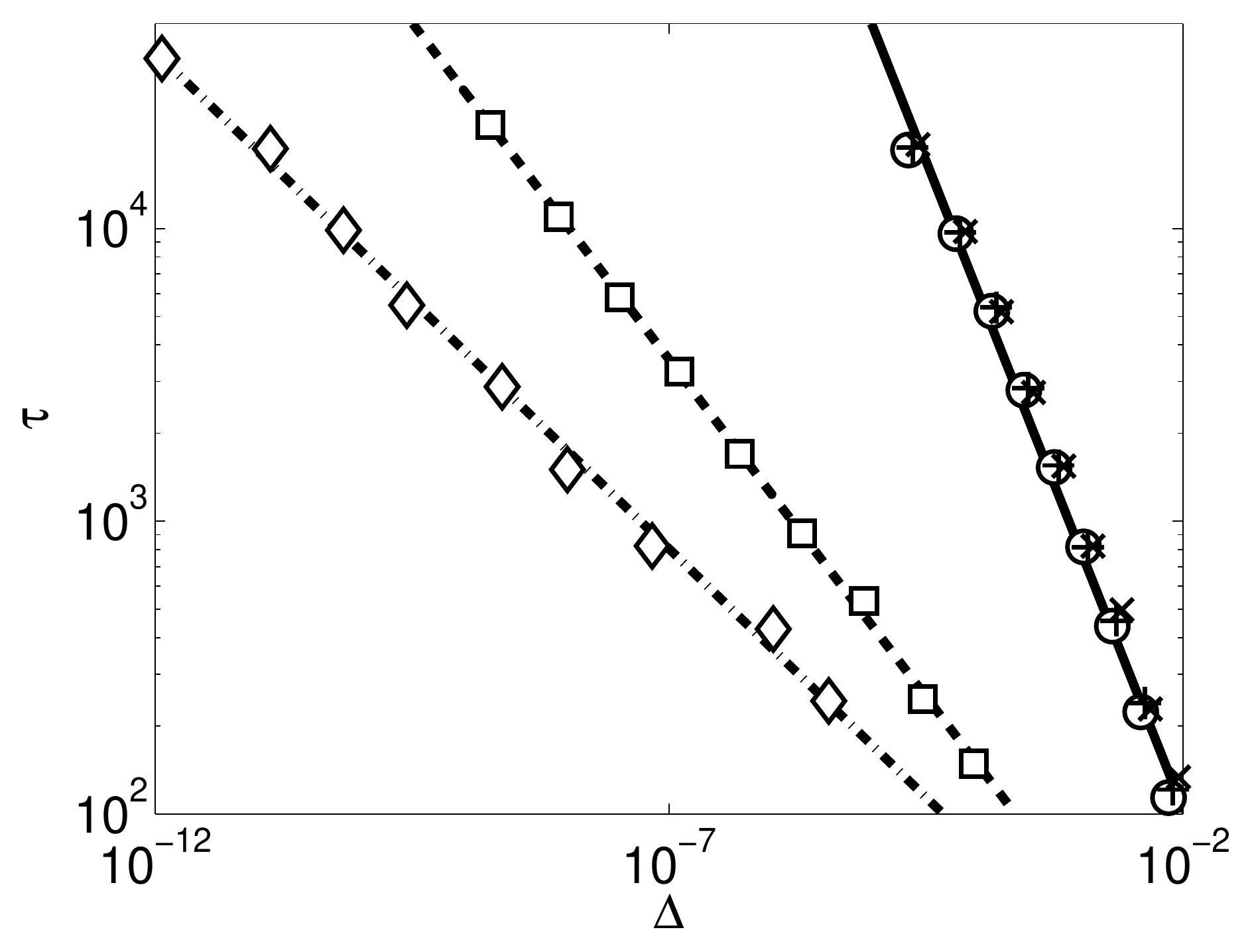}
 \end{center}
 \caption{Computational time for (a) the Fitzhugh-Nagumo, (b) Karma, and (c) Bueno-Orovio models as a function of relative precision $\Delta$. The solid, dashed, dash-dot, and dotted lines are best fits with slopes $\beta=-1$, $\beta=-1/2$, $\beta=-1/3$, and $\beta=-1/4$, respectively. Symbols designate different integration methods, as in Tab.~\ref{tab:conv}. The Bueno-Orovio model (c) includes both the Implicit-Euler and Rush-Larsen methods, partially obscured by the Explicit-Euler results.}
 \label{fig:efficiency}
\end{figure}

The data presented in Figs. \ref{fig:timing} and \ref{fig:convergence} allows us to compare the efficiency of different time-integration methods by computing $\tau$ as a function of $\Delta$. The results are presented in Fig.~\ref{fig:efficiency} and also follow a power law dependence $\Delta\propto \delta t^{\,\beta}$ with scaling exponent $\beta={-1/\alpha}$. Extrapolating the power law fits we discover that, despite its higher complexity and somewhat slower-than-expected convergence rate, the fourth order Runge-Kutta is, by a wide margin, the most efficient time-integration method for all three models considered here for any reasonable relative accuracy (i.e., $\Delta\lesssim 0.01$). For relative accuracy of order $10^{-7}$ required by matrix-free Newton-Krylov method, Runge-Kutta is almost an order of magnitude faster than Heun and three orders of magnitude faster than Euler.

\section{Optimization Directions\label{s:optim}}

Bartocci {\em et al.} \cite{bartocci2011toward} made great improvements to the execution of first-order, single-precision methods, utilizing texture memory on the GPU as a precomputed look-up table for the nonlinear terms of the governing equations. However, texture memory is restricted to single precision values, which makes it of limited use for our present calculations, despite the fast indexation and hardware interpolation features. The utilization of this technique for high-precision integrations would require the implementation of a mixed-precision calculation model, to mitigate the effects of a single-precision texture-memory interpolation.

Currently, our algorithm makes $(n_d + 2) n_s$ function calls per time-step (see Fig. \ref{fig:blockDiagram}). The number of kernel function invocations per time-step can be reduced to $3n_s$ by combining independent diffusion calculations into a single function call. However, the current implementation computes the action of the discrete Laplace operator to minimize initial global memory accesses, by accessing adjacent indices in global memory. Row-adjacent locations in the problem domain are represented by adjacent indices in global memory, and column-adjacent locations are contiguous modulo $N$. As the access pattern becomes more complex (as expected for several variables), the calculation becomes less efficient. As local (n\'ee coalesced) memory access patterns are crucially important for efficient use of the GPU, it is unclear whether a reduction in function calls at the expense of memory access locality would be sufficient to make an integration scheme more efficient.

The algorithm presented in this paper sacrifices efficiency for extensibility: any higher order explicit method of Runge-Kutta type can be implemented by some combination of local, non-local, and Euler-like update kernel functions. Further, reduction of $n_s$ substep methods to $2 n_s$ kernel invocations per time-step is  possible. Non-local terms must be computed separately, but local terms can be combined with the Euler update. This approach yields the smallest number of function calls, corresponding the the number of synchronization points. The odd kernel invocations of the proposed sequence correspond to synchronized non-local computations, and the even invocations to the computation of the local terms and new intermediate state. For example, this would reduce the Runge-Kutta computation of the Karma model from a sequence of twelve functions, to eight.

Our present implementation also does not exploit the vector processing capabilities of the GPU. All OpenCL devices possess a preferred vector processing width for various types of data. The GTX 680 presents a preferred and native vector width for double-precision values as 1; thus there are no gains to be made by writing the integration functions in a vectorized way. On modern CPU hardware (and competing GPU platforms), however, it is not unusual to have a preferred and native vector width of 4 (CPUs) and 2 (AMD's ``Tahiti'' GPUs) for double-precision values, which can improve performance significantly. However, the constraints associated with graphics-oriented nature of symmetric vector processing makes the use of this feature rather nontrivial.

Special functions (natively: sine, cosine, reciprocal, base-2 logarithm and base-2 exponential) are computed separately from multiplication and additon on the GPU, and incur a performance penalty as a result. These functions (and more complicated functions synthesized from these, e.g., the hyperbolic tangent and power function) are sent to the Special Functions Unit (SFU), which computes the result separately from the main thread scheduling unit on the GPU. This asynchrony stems from a base latency for the set of natively computed functions, and longer latencies for the synthesized functions. In this light, the hyperbolic tangent in both the Karma and Bueno-Orovio models should be replaced by $2H_k(x)-1$ with appropriately chosen parameter $k$. Further, our present implementation of the Karma model uses a power function for general values of the parameter $M$. For integer values of this parameter, improvement in speed and accuracy can be achieved by using multiplication instead.

Semi-implicit operator-splitting methods hold a great deal of promise, assuming the Laplacian operator can be inverted cheaply. This requires an efficient implementation of a Fourier transform. Computation of the transform using the Cooley-Tukey algorithm \cite{Cooley65}, requires $\mathcal{O}(N^2\log N^2)$ operations for the forward transform, an additional $\mathcal{O}(2N^2)$ multiplications for the action of the Laplacian in Fourier space, followed by $\mathcal{O}(N^2\log N^2)$ operations for the inverse transform, with $N$ -- the linear dimension of the domain. Finite-difference calculations of the Laplacian requires $\mathcal{O}((2^{s} + 1)N^2)$ operations, where $s$ is the size of the finite-difference stencil. Operator-splitting methods could reduce the overall order of the integration method by reducing the frequency of the calculation of non-local terms, while retaining an accurate approximation of the nonlinearity. Further, sequestering the evaluation of local terms allows one to more efficiently exploit the parallelism of the device, as synchronization need only be enforced before and after the evaluation of the nonlocal terms.

The use of Fourier modes effectively constrains the options for boundary conditions to periodic, compact manifolds. Implementation of Neumann boundary conditions using a mirror-image technique would result in a four-fold increase in the number of grid points, putting the Fourier-based pseudo-spectral method at an even greater disadvantage compared with finite-difference methods. On the other hand, the use of Chebyshev polynomials as a spectral basis affords much greater freedom in the choice of boundary conditions. We should also point out that the numerical cost of computing derivatives is comparable for spectral bases composed of Chebyshev polynomials and Fourier modes \cite{Trefethen00}. Pseudo-spectral methods possess exceptional (exponential) convergence properties, and semi-implicit methods are strongly numerically stable, making their application rather compelling. Hence, we expect the operator-splitting approach to become a serious alternative to finite-difference methods once vendor-agnostic fast Fourier/Chebyshev transform libraries are implemented in OpenCL for the GPU architecture.

\section{Conclusion\label{s:conclus}}

In this paper we have described an implementation of a double precision integrator for simulation of partial-differential models of cardiac dynamics which uses a GPU as a computational accelerator. A hybrid method was used in which a host code written in Matlab executes computational kernels written on OpenCL. We have determined that high-order integration methods, such as the fourth order Runge-Kutta, are substantially faster than lower-order methods regardless of the required accuracy. We have further shown that diffusive coupling through finite-difference stencils is well-suited to highly parallel GPU computations, and allows accurate simulation of the dynamics of two-dimensional models of cardiac dynamics (which are suitable for description of atrial tissue) at speeds approaching real time. For three-dimensional simulations (corresponding to ventricular tissue) it is expected that the GPU performance scaling will be even more dramatic, making intractable calculations on a CPU accessible on a GPU.

\section{Acknowledgements and Funding} 

We would like to thank Ezio Bartocci, Elizabeth Cherry, and Flavio Fenton for their help with developing an OpenCL implementation of these cardiac models and numerous discussions. This material is based upon work supported by the National Science Foundation under Grant No. CMMI-1028133.





\bibliographystyle{model1-num-names} 
\bibliography{../bibtex/cardiac}

\end{document}